\documentstyle[11pt,amsfonts]{article}

\newcommand{\be}{\begin{equation}}
\newcommand{\ee}{\end{equation}}
\newcommand{\bea}{\begin{array}}
\newcommand{\ea}{\end{array}}
\newcommand{\beqa}{\begin{eqnarray}}
\newcommand{\eeqa}{\end{eqnarray}}
\newcommand{\bean}{\begin{eqnarray*}}
\newcommand{\eean}{\end{eqnarray*}}

\def\up#1{\leavevmode \raise.16ex\hbox{#1}}

\newcommand{\gapproxeq}{\lower
 .7ex\hbox{$\;\stackrel{\textstyle >}{\sim}\;$}}
\newcommand{\lapproxeq}{\lower .7ex\hbox{$\;\stackrel
{\textstyle <}{\sim}\;$}}

\newcounter{appendice}

\def\thebibliography#1{{\bf REFERENCES\markboth
 {REFERENCES}{REFERENCES}}\list
 {[\arabic{enumi}]}{\settowidth\labelwidth{[#1]}\leftmargin\labelwidth
 \advance\leftmargin\labelsep
 \usecounter{enumi}}
 \def\newblock{\hskip .11em plus .33em minus -.07em}
 \sloppy
 \sfcode`\.=1000\relax}

\begin{document}
\begin{flushright}
 SU-4252-796\\
\end{flushright}


\centerline{ \LARGE Scale Transformations on the Noncommutative Plane}

\vskip .5cm
\centerline{\LARGE  and the Seiberg-Witten Map } 

\vskip 2cm

\centerline{ {\sc    A. Pinzul$^{a}$ and A. Stern$^{b}$ }  }

\vskip 1cm
\begin{center}
{\it a)  Department of Physics, Syracuse University,\\ Syracuse, 
New York 13244-1130,  USA\\}
{\it b) Department of Physics, University of Alabama,\\
Tuscaloosa, Alabama 35487, USA}

\end{center}

\vskip 2cm

\vspace*{5mm}

\normalsize
\centerline{\bf ABSTRACT} 
We write down three kinds of scale transformations {\tt i-iii)} on the noncommutative
plane.  {\tt i)} is the analogue of standard dilations on the plane,
{\tt ii)} is a re-scaling of the noncommutative parameter $\theta$, and {\tt iii)} is a combination of the previous two, whereby the defining
relations for the noncommutative plane are preserved.  The action of
the three transformations is defined on gauge fields evaluated at  fixed coordinates
and $\theta$.   
 The transformations are obtained only up to
terms which transform covariantly under gauge transformations.  We
give possible constraints on these terms. 
We show how the transformations {\tt i)} and {\tt ii)} depend on the
choice of star product, and show the relation of {\tt ii)} to Seiberg-Witten transformations.  Because {\tt iii)}
preserves the fundamental commutation relations it is a symmetry of
the algebra.  One has 
the possibility of implementing it as a  symmetry of the dynamics, as
well,  in  noncommutative field theories where $\theta$ is
not fixed.

\vspace*{5mm}

\newpage
\scrollmode
\section{Introduction}
Noncommutative field theory is incompatible with conformal field
theory.  Moreover conformal  symmetry is violated in noncommutative theories, at
least in their usual formulation, by
the presence of dimensionfull parameters.  Here we shall be concerned,
in particular, with effects due to dilations.   For the example of the
noncommutative plane,  standard scale transformations of
its coordinates ${\bf x}_i,\; i=1,2$,   do not preserve the defining
commutation relations  \be [{\bf x}_i,{\bf x}_j]
-i\theta\epsilon_{ij}=0\;\label{fndmtlcrs}\;,\ee where $\theta$ is  the
dimensionfull parameter, known as  the 
noncomutativity parameter, and it characterizes the noncommutative
plane.   A re-scaling of $\theta$, corresponding to a mapping
from one noncommutative plane to another, also does not preserve  (\ref{fndmtlcrs}).
Such a re-scaling is generally associated with the Seiberg-Witten map\cite{Seiberg:1999vs}.  On the other hand, we can preserve
(\ref{fndmtlcrs}) with a {\it simultaneous} dilation of the coordinates and a
re-scaling of $\theta$.  Such transformations then define a symmetry
of the algebra.  Moreover, it may also be possibility to implement such  transformations
  as a  symmetry of the dynamics.   
Since the transformations involve a change in $\theta$, as well as
${\tt x}_i$, they act on an ensemble of noncommutative
planes, rather than a single noncommutative plane.   This approach
may  allow one to recover an analogue
of conformal symmetry within the context of noncommutative field
theory.\footnote{For another approach see \cite{Horvathy:2003gi}.}

Concerning the  Seiberg-Witten
map, gauge fields are introduced on the noncommutative plane. Their algebra can
be realized as functions (or symbols) on the commutative plane by working with some
associative star product.  It was shown in \cite{Seiberg:1999vs}
that the symbols ${\cal A}_i$ associated with the noncommutative potentials
could be expressed in terms of commutative potentials ${\cal A}^c_i$,
along with their derivatives, and the noncommutative gauge parameter $\lambda$
could be expressed in terms of the  commutative one $\lambda^c$ and
${\cal A}^c_i$, along with their derivatives.  In Abelian gauge theory the commutative potentials gauge
transform as ${\cal A}^c_i\rightarrow {\cal A}^c_i
+\partial_i\lambda^c $, which then induces a transformation in the
noncommutative potentials ${\cal A}_i({\cal A}^c)\rightarrow {\cal A}_i( {\cal A}^c
+\partial\lambda^c) $ .  In the Seiberg-Witten equations the latter is identified with a
noncommutative gauge transformation of the potentials ${\cal A}_i$.    At first order
in the noncomutativity parameter $\theta$ we don't need to specify the
star product.  So at first order an infinitesimal gauge
variation is given by\beqa \delta^g_\lambda\;{\cal A}_i( {\cal A}^c
)&=&  {\cal A}_i( {\cal A}^c
+\partial\lambda^c)\; -\;{\cal A}_i( {\cal A}^c
)\cr & &\cr & =&\partial_i\lambda (\lambda^c,{\cal A}^c)\;+\;\{
\lambda(\lambda^c,{\cal A}^c),\;{\cal A}_i({\cal A}^c)\} \;,\eeqa where
$\{\;,\;\}$ denotes the Poisson bracket.  For any two functions
${\cal F}$ and ${\cal G}$ on the commutative plane, it is given by
\be \{{\cal F},{\cal G}\} = \theta\;\epsilon_{ij} \partial_i{\cal F}
 \partial_j{\cal G} \;,\label{dfofpb}\ee  where $\partial_i $ are derivatives with
 respect to the coordinates on the plane.
The first order  solutions  for the maps  ${\cal A}_i({\cal A}^c)$ and $\lambda(\lambda^c,{\cal A}^c)$  are 
\beqa {\cal A}_i({\cal A}^c) &=&{\cal A}^c_i \;-\;\frac
\theta 2\epsilon_{jk}\;{\cal A}^c_j
 \;({\cal F}^c_{ik} -\partial_k {\cal A}^c_i)\; +\;
   \frac12 {\cal H}_{{\cal A}^c_i}^{(2)}
\cr & &\cr \lambda( \lambda^c,{\cal A}^c)&=& \lambda^c\; +\;\frac\theta 2 
\;\epsilon_{ij} \; {\cal A}^c_i\;\partial_j
\lambda^c\;\;,\label{fostswes}\eeqa ${\cal F}^c_{ij}$ being the commutative curvature.  We call
${\cal H}^{(2)}_{{\cal A}^c_i} $ a
homogenous term.  It is only required to satisfy
\be {\cal H}^{(2)}_{{\cal A}^c_i+\partial_i\lambda^c}
-{\cal H}^{(2)}_{{\cal A}^c_i} \;=\;\{ \lambda^c,{\cal H}^{(2)}_{{\cal A}^c_i}\}\;,\label{dfofhmtm2}\ee which corresponds to the
first order noncommutative  gauge transformation of a covariant 
field.
Analogous homogeneous terms appear in the  Seiberg-Witten map of
matter fields.  In general the  homogeneous terms  are  undetermined, and such ambiguities in the
construction of 
Seiberg-Witten map are well
known\cite{Asakawa:1999cu},\cite{Jurco:2001rq},\cite{Brace:2001fj},\cite{Barnich:2002pb},\cite{Grimstrup:2003rd},
although they are often ignored in the literature.

Arbitrary homogeneous terms  also result upon making dilations of
the coordinates  ${\bf x}_i$ of the noncommutative plane, as well 
simultaneous dilations and scale transformations in $\theta$.  Here  we find a number of relations connecting the various
homogeneous terms.   We can get additional
 constraints on the homogenous terms  if we demand that 
the gauge fields carry a faithful representation of the two
independent scale transformations, dilations of ${\tt x}_i$ and
re-scalings of $\theta$.  The constraints  allow for  nontrivial
solutions, although the constraints are insufficient in removing all the
ambiguities in the homogeneous terms.  More constraints may result from the presence
of other symmetries, and they may help fix further 
degrees of freedom in the homogeneous terms. 
 Upon generalizing to higher orders in $\theta$, it
becomes necessary to specify the choice of star product, as the answer
 depends on this choice.  We show the explicit dependence of the
 transformations on the choice of star product.

In section 2 we write down the different types of scale
transformations on the noncommutative plane. A fundamental issue is
the construction of operators generating the various transformations.
Concerning
 the generator $D$ of simultaneous dilations and re-scalings of
 $\theta$, we obtain the most general operator that a)  satisfies the Leibniz rule when  acting on the 
product of two functions  on the noncommutative plane $\times\;
{\mathbb{R}}^1$, $
{\mathbb{R}}^1$ parametrized by $\theta$, and b)  annihilates   the left hand
side of (\ref{fndmtlcrs}).   We can then say that $D$ is a generator
of a symmetry of the algebra.   In section 3 we show how gauge fields
transform under these scale transformations.  Our approach closely
follows that of Grimstrup, Jonsson and Thorlacius
\cite{Grimstrup:2003rd}, in that we require the commutator of gauge
transformations with scale transformations to close to gauge
transformations.  This requirement then
insures that all gauge invariant quantities remain
gauge invariant under scale transformations.
As an example, we write down a one parameter family of Seiberg-Witten
maps which interpolate between different star products.   We then write down constraints on the homogeneous terms and  giving some explicit  solutions.
We conclude in section 4 with some preliminary  remarks on the possibility of
implementing simultaneous dilations and re-scalings in $\theta$ as a
symmetry of the dynamics, as well as the algebra.  This  symmetry
is not 
a deformation of the standard dilation symmetry on the plane, but
 rather a new symmetry on the noncommutative plane $\times\;
 {\mathbb{R}}$.

\section{Three Scale Transformations}
\setcounter{equation}{0}

 For simplicity we begin
with the  noncommutative plane at first order in 2.1, and then discuss the
fully noncommutative case in 2.2.

\subsection{Noncommutative Plane at First Order }

The family of noncommutative planes at first order can be defined as
${\mathbb{R}}^3$ modded out by an equivalence relation.
Let ${\mathbb{R}}^3$   be
parametrized by  $(x_1,\;x_2,\;\theta)$.   Then the equivalence relation is
\be\{
x_i,x_j\}-\epsilon_{ij}\; \theta =0 \;,\label{fonp}\ee  and it is the commutative
limit of  (\ref{fndmtlcrs}).
$\{\;,\;\}$ once again  denotes  the
Poisson bracket defined in (\ref{dfofpb}), which is degenerate on ${\mathbb{R}}^3$.

 We consider three
separate scale transformations on  ${\mathbb{R}}^3$ parametrized in
each case by a
real number $\rho$:
\beqa {\tt i)}\quad  (x,\theta) &\rightarrow & (\rho^{-1}
x,\theta)\cr& &\cr
{\tt  ii)}\quad  (x,\theta) &\rightarrow& ( x,\rho^{-2}\theta)\cr & &\cr
{\tt iii)}\quad  (x,\theta) &\rightarrow & (\rho^{-1} x,\rho^{-2}\theta)\label{tstrns}\eeqa
$ {\tt i)}$ is a standard dilation of the  coordinates, ${\tt
  ii)}$ scales $\theta$ and thus  maps to new noncommutative plane,
while ${\tt iii)}$ scales both $x_i$ and $\theta$ in such a way that
it leads to  an automorphism of the algebra defined by (\ref{fonp}).
$ {\tt i)}$ and $ {\tt ii)}$ are independent transformations, while $
{\tt iii)}$ is a combination of $ {\tt i)}$ and $ {\tt ii)}$.

Next introduce representations of these transformations on fields $\phi$ on
${\mathbb{R}}^3$:
\beqa {\tt i)}\quad \phi (x,\theta) &\rightarrow &{}^{\rho_1} \phi (x,\theta)
=e^{\chi_{\rho,{\cal A}}^1}\;\phi (\rho x,\theta)\cr& &\cr
{\tt  ii)}\quad  \phi(x,\theta) &\rightarrow&{}^{\rho_2} \phi (x,\theta)
= e^{\chi_{\rho,{\cal A}}^2}\;\phi( x,\rho^2\theta)\cr& &\cr
{\tt iii)}\quad \phi(x,\theta) &\rightarrow &{}^{\rho_3} \phi (x,\theta)
= e^{\chi_{\rho,{\cal A}}^3}\;\phi(\rho x,\rho^2\theta)\;,\eeqa
where $\chi_{\rho,{\cal A}}^a\;,\;\; a=1,2,3$   are $\rho$ dependent
operators acting  on the space of fields.  The ${\cal A}$ subscript
indicates that representations are, in general, not diagonal, and that
transformations on some field  $\phi$ may involve additional fields ${\cal
  A}_i$.
  ${\tt
  ii)}$ is related to a Seiberg-Witten map.  In the latter,
however, the transformed field is standardly evaluated at the
transformed value of $\theta$, i.e.
\be  {\tt  SW)}\quad  \phi(x,\theta) \rightarrow {}^{\rho_2} \phi
(x,\rho^{-2}\theta) \label{fntsw}
\ee
 Setting  $\rho=1-\epsilon$, $\epsilon$ being infinitesimal, gives the
 infinitesimal version,
\be \phi (x,\theta) \rightarrow  \phi (x,\theta)
+ \delta^{s_a}_\epsilon \phi (x,\theta)  \;,\label{infvrs1to3}\ee
 of
 transformations  {\tt i)-iii)}, where
\beqa {\tt i)} \quad \delta^{s_1}_\epsilon
&=& \epsilon\;(\chi^1_{\cal A} -x_i\partial_i)\;,\qquad
\partial_i=\frac\partial{\partial x_i} \cr & &\cr
{\tt ii})\quad \delta^{s_2}_\epsilon 
&=& \epsilon\;(\chi^2_{\cal A} -2\theta
\partial_\theta)\;,\qquad\partial_\theta=\frac\partial{\partial
  \theta}\cr & &\cr 
{\tt iii)}\quad \delta^{s_3}_\epsilon
&=& \epsilon\;(\chi^3_{\cal A} -D)
\;,\qquad  D=x_i\partial_i +2\theta
\partial_\theta\;,\label{infscltrns}\eeqa with
$\chi^a_{\cal A}=\lim_{\epsilon\rightarrow 0}\frac{ \chi^a_{\rho,{\cal
      A}}} \epsilon $. The infinitesimal variations $\delta^{s_a}_\epsilon$ are related by
$ \delta^{s_3}_\epsilon=\delta^{s_1}_\epsilon+\delta^{s_2}_\epsilon  $, 
leading to the constraint on $\chi^a_{\cal A}$ 
\be\chi^3_{\cal A}=\chi^1_{\cal A}+\chi^2_{\cal A}\label{cnstonch}\ee

 When acting on the Poisson bracket of two
functions ${\cal F}$ and ${\cal G}$ on
${\mathbb{R}}^3$,
 $D$ satisfies the Leibniz rule
\be D\{{\cal F},{\cal G}\}(x,\theta)= \{D{\cal F},{\cal G}\}(x,\theta)
+ \{{\cal F},D{\cal G}\}(x,\theta)\;\label{lbnzrl} \ee Variations
$\delta^{s_a}_\epsilon$ also satisfy the Leibniz rule as they are
evaluated at   fixed coordinates in  ${\mathbb{R}}^3$.     Using
(\ref{lbnzrl}), $D$ is seen to annihilate the left hand side of
(\ref{fonp}), and so transformations  ${\tt iii)}$ leave invariant the
equivalence relation.
On the other hand, $x_i\partial_i$ and $2\theta \partial_\theta$ do
not satisfy the Leibniz rule, but rather: \beqa x_i\partial_i\{{\cal F},{\cal G}\}(x,\theta)&=& \{x_i\partial_i{\cal F},{\cal G}\}(x,\theta)
+ \{{\cal F},x_i\partial_i{\cal G}\}(x,\theta) -2\{{\cal F},{\cal
  G}\}(x,\theta)\label{vlnlr1} \\ & &\cr
 2\theta\partial_\theta\{{\cal F},{\cal G}\}(x,\theta)&=& \{2\theta\partial_\theta{\cal F},{\cal G}\}(x,\theta)
+ \{{\cal F},2\theta\partial_\theta{\cal G}\}(x,\theta) +2\{{\cal F},{\cal
  G}\}(x,\theta)\; \label{nolbnzrl}
 \eeqa  Note that with these modified product rules, $x_i\partial_i$ and $2\theta \partial_\theta$ also  annihilate the left hand side of
(\ref{fonp}), and hence  leave invariant the
equivalence relation.

The infinitesimal version of the Seiberg-Witten map  (\ref{fntsw})
is \be {\tt SW)}\quad \phi (x,\theta) \rightarrow  \phi (x,\theta)
+ \delta^{SW} \phi(x,\theta)  \;,\qquad \delta^{SW} =\delta^{s_2}_\epsilon + 2\epsilon\theta
\partial_\theta\;,\ee and so this variation is given solely  by
$\chi^2_{\cal A}\;$, \be \delta^{SW}= \frac
{\delta\theta}{2\theta}\;\chi^2_{\cal A}\;, \label{infsw} \ee
where $\delta\theta = 2\epsilon\theta$.
Acting on the Poisson bracket it then does not  satisfy the Leibniz
rule, but rather \be \delta^{SW}\{{\cal F},{\cal G}\} = \{\delta^{SW}{\cal F},{\cal
  G}\}+\{{\cal F},\delta^{SW}{\cal
  G}\}+\frac{\delta\theta}\theta\{{\cal F},{\cal
  G}\}\label{swtafolrv}\ee   In the next subsection we show how this
result generalizes in the full noncommutative theory.  In that case, the
Seiberg-Witten variation $\delta^{SW}$ acting on a product of functions violates
the Leibniz rule.  This is since the variation $\delta^{SW}$ compares
functions at different values of $\theta$.

\subsection{Noncommutative Plane  to All Orders }

To go to all orders we replace $x_i$ by operators   ${\bf x}_i$,
$i=1,2$.  The latter
satisfy  (\ref{fndmtlcrs}) and  generate the associative algebra corresponding to the
two-dimensional    Moyal (or noncommutative) plane.  The Moyal plane is characterized by
$\theta$, which remains a c-number.   We consider the analogue of the three
 scale transformations (\ref{tstrns}).
$ {\tt i)}$ is now a dilation of operators   ${\bf x}_i$, ${\tt
  ii)}$  maps between different Moyal planes, and ${\tt iii)}$ scales
both ${\bf x}_i$ and $\theta$ in such a way that
it leads to  an automorphism of the algebra  (\ref{fndmtlcrs}).

Concerning $ {\tt i)}$, the analogue of the dilation generator
$x_i\partial_i$ is ambiguous.   $\frac12 [{\bf x}_i,\nabla_i
\quad]_+$ was suggested in \cite{Grimstrup:2003rd},  where
$[\;,\;]_+$ denotes the anticommutator and  $\nabla_i$ is the inner
derivative on the noncommuting plane.  Acting on some function $F$ the
latter is given by
\be  \nabla_i F =\frac i\theta\epsilon_{ij}[{\bf x}_j,F]\ee  For a more
general noncommutative dilation generator, we  add $-2\theta\;\tau$ to  $\frac12 [{\bf x}_i,\nabla_i
\quad]_+$, where $\tau$ is  a linear operator
acting 
   on the space of functions on the noncommuting plane.  Below we will
   obtain several constraints on $\tau$.

Concerning ${\tt ii)}$ we need to define an analogue of the derivative
$\partial_\theta$.  We call it $\nabla_\theta$.  Unlike  $\nabla_i$,
it is not an inner
derivative.  We instead define $\nabla_\theta$ such that it commutes
with $\nabla_i$ and it is the ordinary
derivative on a c-number valued function $f$ of $\theta$,
i.e. $\nabla_\theta f(\theta) =\partial_\theta f(\theta) $.  We shall
also require that it satisfies a  product rule such that  $\nabla_\theta$  annihilates   the left hand
side of (\ref{fndmtlcrs}), and it is  consistent  with
the associativity of the algebra, i.e.  $\nabla_\theta ( (FG)H) =
\nabla_\theta ( F(GH))$.  This product rule is not the Leibniz rule.

With
regard to ${\tt iii)}$ we will need to construct the noncommutative
analogue of the derivative operator $D$, which for convenience we also call
$D$.  We define it, as in the case of first order noncommutativity, to
be the sum of the generators for  ${\tt
  i)}$ and  ${\tt
  ii)}$.  Thus acting on function $F$  on the noncommutative plane $\times\;
{\mathbb{R}}^1$ ( ${\mathbb{R}}^1$ being parametrized by $\theta$),  \be DF =\frac12 [{\bf
  x}_i,\nabla_iF]_+ +2\theta (\nabla_\theta F -\tau(F)\;)\;\label{fncD}
\ee   In order to recover the first order result, we need that  $\tau$
acting on fields vanishes
   at lowest order in $\theta$.
 For the fully noncommutative $D$ we shall require that  a), unlike  $\nabla_\theta$, it  satisfies the Leibniz rule when  acting on the 
product of two functions $F$ and $G$ on the noncommutative plane $\times\;
{\mathbb{R}}^1$ \be  D (FG) =( D F)G+ F(D G)\label{lrfdmp}\; \ee  
and b) it  annihilates   the left hand
side of (\ref{fndmtlcrs}) when evaluated on the noncommutative plane.
For this we need that
\be \tau([{\bf x}_i,{\bf x}_j]) =[\tau({\bf x}_i),{\bf x}_j] +[{\bf
  x}_i,\tau({\bf x}_j)]\label{ctxx} \ee  Later in this section we find it
convenient to  impose  stronger conditions on $\tau$:
\be \tau({\tt x}_i)\; \propto\; {\rm central}\;{\rm
  element}\;,\qquad \tau( {\rm central}\;{\rm
  element})=0 \label{txice} \ee
From a) $D$ is consistent with the
associativity of the algebra, and also agrees with (\ref{lbnzrl}) at
first order.  From b) $D$  generates a symmetry of the
algebra.

Next define fields $\Phi$ belonging to a bimodule on the noncommutative plane $\times\;
{\mathbb{R}}^1$, with
an associative product.
For the analogue of the infinitesimal variations  (\ref{infvrs1to3}) and
  (\ref{infscltrns}) we take 
\be \Phi ({\bf x},\theta) \rightarrow  \Phi ({\bf x},\theta)
+ \delta^{s_a}_\epsilon \Phi ({\bf x},\theta)  \;,\ee
\beqa {\tt i)} \quad \delta^{s_1}_\epsilon \Phi
&=& \epsilon\;(\chi^1_{ A} \Phi -\frac12 [{\bf x}_i,\nabla_i
\Phi]_++2\theta\;\tau(\Phi)\;)\;, \cr & &\cr
{\tt ii})\quad \delta^{s_2}_\epsilon  \Phi
&=& \epsilon\;(\chi^2_{ A}\Phi
  -2\theta
\nabla_\theta\Phi )\;,
\cr & &\cr 
{\tt iii)}\quad \delta^{s_3}_\epsilon\Phi 
&=& \epsilon\;(\chi^3_{ A}\Phi -D\Phi)\;\label{fncdfvrs}\eeqa
   As in the previous subsection we require  $
\delta^{s_3}_\epsilon=\delta^{s_1}_\epsilon+\delta^{s_2}_\epsilon  $ and
hence we have the analogue of (\ref{cnstonch}).

We saw in  (\ref{vlnlr1}) that $\partial_\theta$ does not satisfy the Leibniz
 rule when acting on the Poisson bracket.  Similarly,  $\nabla_\theta$ 
 satisfies a modified product rule in the fully noncommutative theory.
 The modified product rule for  $\nabla_\theta$  should be consistent with (\ref{lrfdmp}),
 as well as (\ref{vlnlr1}) at first order.
We first obtain it for the case $\tau=0$ and  then generalize to arbitrary $\tau$.

\begin{enumerate}
\item $\tau=0$.  To determine the
product rule for  $\nabla_\theta$
 we first write down the product rule for dilations 
\be  \frac12 [{\bf x}_i,\nabla_i (FG)]_+ =  \frac12 [{\bf
  x}_i,\nabla_i F]_+G+ \frac12 F [{\bf x}_i,\nabla_i G]_+
-i\theta\epsilon_{ij} \nabla_iF\nabla_jG\;\ee
We note that with this product rule, dilations  annihilate   the left hand
side of (\ref{fndmtlcrs}).  For consistency with (\ref{lrfdmp}) we
need  the following
product rule for  $\nabla_\theta$:
\be   \nabla_\theta ( FG) =  (\nabla_\theta F)G+
 F(\nabla_\theta G)+\frac i{2}\epsilon_{ij} \nabla_iF\nabla_jG \label{vltnolr}
 \ee 
 With this product rule $\nabla_\theta$  annihilates   the left hand
side of (\ref{fndmtlcrs}).  Furthermore, it is easily seen that the product rule (\ref{vltnolr}) is consistent  with
the associativity of the algebra.  This also follows from the fact
that    (\ref{vltnolr}) agrees with the  Moyal-Weyl star product realization of the
operator algebra\cite{groe}.   For this functions $F$, $G$,...
on the noncommuting plane  $\times\;
{\mathbb{R}}^1$ are replaced by
symbols ${\cal F}_0$, ${\cal G}_0$,... in the Moyal-Weyl star product representation, which are functions on
the commuting plane  $\times\;
{\mathbb{R}}^1$.  As the star product depends on
   $\theta$, $\nabla_\theta$ acts nontrivially on the star product, in the
   sense that the Leibniz rule is not satisfied.  The
   Moyal-Weyl star, which we denote by $\star_0$, acting on any two  symbols is given by
\be \star_0 = \exp\;\biggl\{ \frac {i\theta}2 \epsilon_{ij}\overleftarrow{
  \partial_i}\;\overrightarrow{ \partial_j} \biggr\}  \ee  $\overleftarrow{
  \partial_i}$ and $\overrightarrow{ \partial_j}$ are left and right
derivatives on the commuting plane, respectively.   Acting with
$\nabla_\theta$ gives
\be \nabla_\theta\;\star_0 =  \frac i2 \; \epsilon_{ij}\overleftarrow{
  \partial_i}\;\overrightarrow{ \partial_j}\; \;\star_0 \ee   Then for
any two functions ${\cal F}_0$ and ${\cal G}_0$ on the plane
\be \nabla_\theta\; ( {\cal F}_0\star_0{\cal G}_0) =   \nabla_\theta{\cal
    F}_0\star_0 {\cal G}_0 + {\cal F}_0 \star_0  \nabla_\theta{\cal
    G}_0   + \frac i2 \; \epsilon_{ij}
  \partial_i{\cal F}_0\star_0 { \partial_j}{\cal G}_0  \label{dtosp}\;,
  \ee which agrees with
(\ref{vltnolr}). 

\item Arbitrary $\tau$.  Now in order to recover (\ref{lrfdmp}), the
  product rule for $\nabla_\theta$ should be changed to \be   \nabla_\theta ( FG) =  (\nabla_\theta F)G+
 F(\nabla_\theta G)+\frac i{2}\epsilon_{ij} \nabla_iF\nabla_jG
 +\tau(FG) - \tau(F)G - F\tau(G)\label{mstgnrldlt}\ee   The extra terms are consistent
 with the associativity of the product, and using (\ref{ctxx})
 $\nabla_\theta$ again  annihilates   the left hand
side of (\ref{fndmtlcrs}).  
With this  choice $D$  acting on the
product of two fields again
satisfies the Leibniz rule.  
 The choice of (\ref{mstgnrldlt}) can also be motivated by
a star product on the noncommuting plane, only now it is not the
  Moyal-Weyl star product.   Rather,  it is a
star product which can be  obtained by a general Kontsevich map $T$  from the
Moyal-Weyl star product.
  $T$ is a nonsingular operator which maps any pair of symbols  ${\cal F}_0$ and ${\cal
  G}_0$ in the Moyal-Weyl star product representation to a new pair of symbols  ${\cal F}$ and ${\cal G}$, which
realize the noncommutative algebra with respect to the new 
star product, which we denote by $\star$, with the fundamental
  property 
\be  {\cal F}\star{\cal G} = T( {\cal F}_0\star_0{\cal
  G}_0)\;,\qquad{\cal F}= T( {\cal F}_0)\;,\qquad  {\cal G}= T( {\cal G}_0) \ee
  Variations 
in $\theta$ of a symbol ${\cal F}$ can be expressed as 
\be \delta {\cal F}= d\theta\; \nabla_\theta {\cal F}=\delta T ( {\cal
  F}_0) + T (\delta {\cal F}_0 ) =d\theta \;t ({\cal
  F} ) + T(\delta{\cal F}_0   )\;, \label{delcalf} \ee where $t= \frac{\delta
  T}{\delta\theta}T^{-1}$.
Applying this to the star product of  ${\cal F}$ and ${\cal G}$ and using (\ref{dtosp}) gives
\beqa \delta ({\cal F\star {\cal G}}) & =&d\theta \;t ({\cal
  F}\star{\cal G} ) + T(\;\delta({\cal F}_0 \star_0{\cal G}_0)\;
)\\ & &\cr & =&d\theta \;t ({\cal
  F}\star{\cal G} ) + T\biggl(\; \delta{\cal
    F}_0\star_0 {\cal G}_0 + {\cal F}_0 \star_0  \delta{\cal
    G}_0   + \frac i2 d\theta\; \epsilon_{ij}
  \partial_i{\cal F}_0\star_0 { \partial_j}{\cal G}_0 
\;
\biggr)\cr& &\cr & =&d\theta \;t ({\cal
  F}\star{\cal G} ) + T( \delta{\cal
    F}_0)\star {\cal G} + {\cal F} \star T( \delta{\cal
    G}_0 )  + \frac i2 d\theta\; \epsilon_{ij}
 T( \partial_i{\cal F}_0)\star T( { \partial_j}{\cal G}_0)\nonumber  \eeqa
Finally using (\ref{delcalf}) and assuming that $T$ commutes with
$\partial_i$,
\be \delta ({\cal F\star {\cal G}})  =\delta{\cal
    F}\star {\cal G} + {\cal F} \star  \delta{\cal
    G} +  
d\theta \;\biggl(t ({\cal
  F}\star{\cal G} ) - t({\cal
    F})\star {\cal G} - {\cal F} \star t( {\cal
    G} )  + \frac i2  \epsilon_{ij}
 \partial_i {\cal F} \star  \partial_j{\cal G}\biggr)\;,  \ee 
and then
\be \nabla_\theta ({\cal F\star {\cal G}})  =\nabla_\theta{\cal
    F}\star {\cal G} + {\cal F} \star  \nabla_\theta{\cal
    G} +  
t ({\cal
  F}\star{\cal G} ) - t({\cal
    F})\star {\cal G} - {\cal F} \star t( {\cal
    G} )  + \frac i2  \epsilon_{ij}
 \partial_i {\cal F} \star  \partial_j{\cal G}  \ee 
 This
 result  agrees with (\ref{mstgnrldlt}) upon interpreting $t({\cal F})$
 and $t({\cal G})$ as  the symbols of $\tau (F)$ and $\tau (G)$,
 respectively, with the product of functions realized by the $\star$.
 The only assumption used in the above was that $T$ commutes with
$\partial_i$, or equivalently
\be [\nabla_i,\tau]= 0\label{nbltcmt}\ee   This condition implies
(\ref{txice}).   To prove this let  $[\nabla_i,\tau]$ act on ${\tt x}_j$.  If $i\ne j$, then  
(\ref{nbltcmt}) implies $\nabla_i\;\tau({\tt x}_j)= 0$, or $\tau({\tt
  x}_j)$ is a function of only ${\tt x}_j$.  For $i=j$, (\ref{nbltcmt})
gives \be\tau\biggl(\frac i\theta \epsilon_{ik}[{\tt x}_k,{\tt x}_i]\biggr) = 
\frac i\theta \epsilon_{ik}[{\tt x}_k,\tau({\tt x}_i)]\;, \qquad{\rm
  no}\;{\rm sum}\;{\rm on} \;i,\ee or  $\tau([{\tt x}_k,{\tt x}_i]) = 
[{\tt x}_k,\tau({\tt x}_i)]$, with $i\ne k$.  Then from (\ref{ctxx}),
$[\tau({\tt x}_k),{\tt x}_i] = 0$,  and so the only possibility for  $\tau({\tt
  x}_k)$ is given in  (\ref{txice}).  $\tau(\theta)=0$ then follows
from (\ref{ctxx}).

\end{enumerate}

At all orders in $\theta$,
 infinitesimal 
 Seiberg-Witten transformations are given by 
 \be {\tt SW)}\quad \Phi \rightarrow  \Phi 
+ \delta^{SW} \Phi  \;,\qquad \delta^{SW} =\delta^{s_2}_\epsilon + 2\epsilon\theta
\nabla_\theta\;\ee  The variation $\delta^{SW} $ is thus determined by $\chi_A^2$,
 as in (\ref{infsw}).   $\chi_A^2$ acting on a product of functions does not respect the Leibniz
rule, and hence neither does  $\delta^{SW} $.  For the
 case of arbitrary $\tau$,
\be  \delta^{SW} ( FG) = ( \delta^{SW} F)G+
 F(\delta^{SW} G)+\delta\theta\biggl(\frac i{2}\epsilon_{ij} \nabla_iF\nabla_jG
 +\tau(FG) - \tau(F)G - F\tau(G)\biggr)\;,\ee which agrees with
 (\ref{swtafolrv}) at first order.

We end this section by giving  an
 example of a nonvanishing $\tau$, or  actually a one parameter family
 of operators $\tau_\beta$, where $\beta$ is a real parameter. 
Consider first a one parameter family of  Kontsevich maps $ T=T_\beta$ given by
\be T_\beta=\exp\biggl\{ \frac{\beta\theta}4 \partial_i\partial_i\biggr\}\label{opfkms} \ee
   The corresponding family 
of operators is $\tau_\beta =\frac{\beta}4\;\nabla_i\nabla_i$.
Here $\tau_\beta({\tt x}_i)=0$, and so condition  (\ref{txice}) is
 satisfied.\footnote{ Concerning the requirement that linear operators $\tau$
 vanish at lowest order in $\theta$, we show in section 3.3 that
 this is the case for
 $\tau_\beta$ acting on  fields belonging to nontrivial representations of
 the gauge group,
 up to a gauge transformation and  covariant terms, the latter of which can be
 absorbed in the homogeneous terms.}  Substituting into
(\ref{mstgnrldlt}) gives a one parameter family of product rules
\be   \nabla_\theta ( FG) = ( \nabla_\theta F)G+
 F(\nabla_\theta G)+\frac i{2}\epsilon_{ij} \nabla_iF\nabla_jG
 + \frac\beta 2 \nabla_i F\nabla_i G  \label{opvltnolr}\ee  The one parameter family of product rules
   corresponds to a one parameter family of star products,
\be \star_\beta = \exp\;\biggl\{ \frac {\theta}2\;\bigl[i \epsilon_{ij}\overleftarrow{
  \partial_i}\;\overrightarrow{ \partial_j}+\beta\;\overleftarrow{
  \partial_i}\;\overrightarrow{ \partial_i} \bigr] \biggr\}\label{opfosps}  \ee
  The case of $\beta=1$ corresponds to the Voros star
 product\cite{vor},\cite{Zachos:1999mp},\cite{Alexanian:2000uz} $\star_1$,
which can be written \be\star_1 =
\exp{\biggl\{\theta \;\overleftarrow{ \frac\partial{ \partial\zeta }}\;\;
\overrightarrow{ {\partial\over{ \partial\bar\zeta} }}\biggr\}}\;,\qquad
\zeta = \frac{x_1+ix_2}{\sqrt{2}}\;,\quad\bar \zeta = \frac{x_1-ix_2}{\sqrt{2}}\ee
Thus $\star_\beta$ is a one-parameter family of star products
connecting the Voros and  Moyal Weyl products.

\section{Gauge Transformations}
\setcounter{equation}{0}

Next we introduce  $U(1)$ gauge
theory   in the fully noncommutative theory.   $\chi^a_{ A}$
can be  determined for any given star product, up to homogeneous terms,  by
demanding that the commutator of transformations (\ref{fncdfvrs}) with gauge transformations is also a
gauge transformation.\footnote{Actually, when obtaining the
  Seiberg-Witten map it is common  to demand the stronger condition that the
  commutator vanishes .} \cite{Grimstrup:2003rd}   This requirement then
insures that all gauge invariant quantities remain
gauge invariant under scale transformations ${\tt i-iii)}$.

 In 3.1 below we  consider the case of fields $\Phi$ which are  covariant under gauge
transformations.  This means that  infinitesimal gauge transformations parametrized
by  $\Lambda$ (an
  infinitesimal function of the noncommutative plane $\times\;
{\mathbb{R}}^1$)  of  $\Phi$
  are given by
\be\Phi \rightarrow \Phi^\Lambda =\Phi + \delta^g_\Lambda \Phi\;,\qquad
\delta^g_\Lambda \Phi = i[\Lambda,\Phi]\;,\label{gtsop}\ee
Potentials $A_i$ and field strength $F_{ij}$ are considered in 3.2.
 $\chi_{ A}^a$ depends in general on potentials, as the notation implies.  Under gauge
 transformations
\be A_i \rightarrow A_i^\Lambda =A_i + \delta^g_\Lambda A_i\;,\qquad
\delta^g_\Lambda A_i = D_i\Lambda\equiv i[\Lambda,A_i-\epsilon_{ij}\frac{{\bf  x}_j}\theta
]\;\label{gtsoAi}\;,\ee while $F_{ij}$ transforms covariantly.  In 3.3
we give a one parameter family of Seiberg-Witten transformations,
while in 3.4 we derive some constraints on the homogeneous
contributions to the solutions.

\subsection{Covariant fields}

In  computing the   commutator of transformations (\ref{fncdfvrs})
with gauge transformations on a covariant field $\Phi$, we
first consider 
 the simpler case of $\tau=0$, and then the case of arbitrary $\tau$.
\begin{enumerate}
\item For the case  $\tau=0$ we need the product rule (\ref{vltnolr}).
  The commutators are
\beqa  [\delta^{s_1}_\epsilon, \delta^g_\Lambda]\;\Phi &=&\epsilon
\;(-\delta^g_\Lambda\chi^1_{{ A}}\Phi+i[\Lambda,\chi^1_{   A}\Phi]
+i[\chi^1_{{ A}}\Lambda
,\Phi]+\theta\epsilon_{ij}\;[\nabla_i\Lambda,\nabla_j\Phi]_+\;)  \cr &
&\cr[\delta^{s_2}_\epsilon, \delta^g_\Lambda]\;\Phi &=&\epsilon \;(-\delta^g_\Lambda\chi^2_{{ A}}\Phi+i[\Lambda,\chi^2_{   A}\Phi]
+i[\chi^2_{{ A}}\Lambda
,\Phi]-\theta\epsilon_{ij}\;[\nabla_i\Lambda,\nabla_j\Phi]_+\;)
\cr & &\cr [\delta^{s_3}_\epsilon, \delta^g_\Lambda]\;\Phi &=&\epsilon
 \;(-\delta^g_\Lambda\chi^3_{{ A}}\Phi+i[\Lambda,\chi^3_{   A}\Phi]
+i[\chi^3_{{ A}}\Lambda ,\Phi]\;)\label{thrcmtrssagfnc}
\eeqa 
These are equal to gauge variations, i.e.
\be [\delta^{s_a}_\epsilon, \delta^g_\Lambda]\;\Phi =
\delta^g_{\Lambda^{(a)}_\epsilon}\Phi\;,\ee when
\beqa  
\chi^1_{A}\Phi&=& \;\; \frac\theta 2\epsilon_{ij} [ A_i,
D_j\Phi +\nabla_j\Phi ]_+ + H^{(1)}_\Phi\label{fncstcone}\\  & &\cr
\chi^2_{A}\Phi&=&- \frac\theta 2\epsilon_{ij} [ A_i,
D_j\Phi +\nabla_j\Phi ]_+ + H^{(2)}_\Phi\label{fncstctwo}\\& &\cr
\chi^3_{A}\Phi&=&\qquad\qquad\qquad\qquad\qquad\;\;\quad H^{(3)}_\Phi\;,\label{fncchiaoph}
 \eeqa  where the  covariant derivative is defined in (\ref{gtsoAi})
 and  $H^{(a)}_\Phi$
 are defined to transform
 covariantly under gauge transformations, i.e.
\be\delta^g_\Lambda H^{(a)}_\Phi=i[\Lambda,H^{(a)}_\Phi]\ee 
So  possible solutions are   $H^{(a)}_\Phi$ proportional  to $\Phi$.   More
generally,   $H^{(a)}_\Phi$ can be any polynomial of covariant fields and
 their covariant derivatives.  We
 call these  homogeneous solutions. From (\ref{cnstonch}) and
(\ref{fncstcone}-\ref{fncchiaoph}),  the homogenous solutions are
related by $  H^{(3)}_\Phi=
H^{(1)}_\Phi+ H^{(2)}_\Phi$.    If  $ H^{(3)}_\Phi$ vanishes, then $\chi^3_{
   A}$ annihilates $\Phi$.   If  $ H^{(3)}_\Phi$ is proportional to $\Phi$ it
 is an eigenvector  of $\chi^3_{   A}$.
The gauge parameters $\Lambda^{(a)}_\epsilon$ corresponding to
(\ref{fncstcone}), (\ref{fncstctwo}) and (\ref{fncchiaoph}) are given by
\beqa  
 \Lambda_\epsilon^{(1)}& =& \epsilon\;\biggl(\chi^1_{ A}\Lambda+\frac\theta
 2\epsilon_{ij} [A_i,\nabla_j \Lambda]_+ \biggr)\label{lonee}\\   & &\cr
 \Lambda_\epsilon^{(2)}& =& \epsilon\;\biggl(\chi^2_{ A}\Lambda-\frac\theta 2\epsilon_{ij} [A_i,\nabla_j \Lambda]_+ \biggr)\label{ltwee}\\& &\cr
 \Lambda_\epsilon^{(3)}& =& \epsilon\;\chi^3_{ A}\Lambda\;\label{lmde}\;,\eeqa
 respectively.   Had we imposed the stronger condition that scale
 transformations commute with gauge transformations, then $
 \Lambda_\epsilon^{(a)}=0$, and we would obtain unambiguous  solutions
 for $\chi^a_{ A}\Lambda$, and hence
 $\delta^{s_a}_\epsilon \Lambda$.

\item  When $\tau\ne 0\;$,  $\chi_A^1$ gets replaced by
  $\chi_A^1+2\theta\;\tau$, and so (\ref{fncstcone}) and (\ref{lonee})
  are  generalized to
\beqa\chi^1_{A}\Phi&=& \frac\theta 2\epsilon_{ij} [ A_i,
D_j\Phi +\nabla_j\Phi ]_+-2\theta\tau(\Phi) + H^{(1)}_\Phi\;, \cr & &\cr
 \Lambda_\epsilon^{(1)}& =& \epsilon\;\biggl(\chi^1_{ A}\Lambda+2\theta\tau(\Lambda)+\frac\theta
 2\epsilon_{ij} [A_i,\nabla_j \Lambda]_+ \biggr)\label{fncc1wt}\eeqa Concerning
 $\chi_A^2\;$, the commutator $ [\delta^{s_2}_\epsilon,
 \delta^g_\Lambda]\;\Phi$ in (\ref{thrcmtrssagfnc}) picks up the additional
 terms $$2i\epsilon \theta\; \biggl(\tau([\Lambda,\Phi]) -
 [\tau(\Lambda),\Phi]-[\Lambda,\tau(\Phi)]\biggr)\;,$$  and one can
 write the answer by replacing  $\chi_A^2\;$ by
 $\chi_A^2-2\theta\;\tau$ in the previous result.  As a result
 (\ref{fncstctwo}) and (\ref{ltwee}) are
 generalized to \beqa\chi^2_{A}\Phi&=& -\frac\theta 2\epsilon_{ij} [ A_i,
D_j\Phi +\nabla_j\Phi ]_++2\theta\tau(\Phi) + H^{(2)}_\Phi\;,\cr & &\cr
 \Lambda_\epsilon^{(2)}& =& \epsilon\;\biggl(\chi^2_{ A}\Lambda-2\theta\tau(\Lambda)-\frac\theta
 2\epsilon_{ij} [A_i,\nabla_j \Lambda]_+ \biggr)\label{fncc2wt}\eeqa
So up to the homogeneous terms, $\chi^1_{A}\Phi$ and $\chi^2_{A}\Phi$
only differ by a sign.
In comparing (\ref{fncc1wt}) and   (\ref{fncc2wt}) it follows that the
sum, i.e.
$\chi^3_A\Phi$ and the corresponding gauge parameter $\Lambda^{(3)}_\epsilon$, is unaffected by the generalization, and thus still given by 
(\ref{fncchiaoph}) and (\ref{lmde}), respectively.

From now on we consider the most general case where $\tau$ is not
necessarily zero. 
\end{enumerate}

   From  (\ref{fncc2wt}), using (\ref{infsw}),  we get  the general
expression for the Seiberg-Witten variation of $\Phi$ from $\chi^2_A$.
  The term $\tau(\Phi)$
depends on the choice of the star product (for example, it vanishes in the case of
the Moyal-Weyl star), while  $ H^{(2)}_\Phi$, represents the homogeneous
contributions.  At lowest order in
$\theta$,  the result agrees with the solutions (\ref{fostswes}).

\subsection{Potentials and field strength}

We next determine the variations $\delta^{s_a}_\epsilon$  of the
gauge   potentials and field strength. The latter is given by 
\be F_{ij} =\nabla_i A_j-\nabla_j A_i - i[A_i,A_j]\ee
  Concerning the potentials,
\beqa {\tt i)} \quad \delta^{s_1}_\epsilon A_i
&=& \epsilon\;\biggl(\chi^1_{ A} A_i -\frac12 [{\bf x}_j,\nabla_j
A_i]_++2\theta\;\tau(A_i)\biggr)\; \cr & &\cr
{\tt ii})\quad \delta^{s_2}_\epsilon  A_i
&=& \epsilon\;(\chi^2_{ A}A_i
  -2\theta
\nabla_\theta A_i )\;
\cr & &\cr 
{\tt iii)}\quad \delta^{s_3}_\epsilon A_i 
&=& \epsilon\;(\chi^3_{ A}A_i -DA_i)\;\label{fncdfvrsai}\eeqa
  We thus need to determine $\chi^{a}_A {
   A}_i$.   
For this we can first look at variations $\delta^{s_a}_\epsilon$  of the
 covariant
 derivative of $\Phi$\beqa {\tt i)} \quad \delta^{s_1}_\epsilon D_i\Phi
&=&  D_i  \delta^{s_1}_\epsilon \Phi  + i[\Phi,  \delta^{s_1}_\epsilon
A_i ] =
   \epsilon\;\biggl(\chi^1_{ A} D_i\Phi -\frac12 [{\bf x}_j,\nabla_j
D_i\Phi]_++2\theta\;\tau(D_i\Phi)\biggr)\; \cr & &\cr
{\tt ii})\quad \delta^{s_2}_\epsilon  D_i\Phi
&=& D_i  \delta^{s_2}_\epsilon \Phi  + i[\Phi,  \delta^{s_2}_\epsilon
A_i ] = \epsilon\;(\chi^2_{ A}D_i\Phi
  -2\theta
\nabla_\theta D_i\Phi )\;
\cr & &\cr 
{\tt iii)}\quad \delta^{s_3}_\epsilon D_i\Phi 
&=& D_i  \delta^{s_3}_\epsilon \Phi  + i[\Phi,  \delta^{s_3}_\epsilon
A_i ] = \epsilon\;(\chi^3_{ A}D_i\Phi -DD_i\Phi)
\label{fncdfvrsdphi}\eeqa  Substituting (\ref{fncdfvrs}) and
(\ref{fncdfvrsai}) gives
\beqa
[D_i,\chi^1_A +2\theta\; \tau]\;\Phi &=&\nabla_i\Phi -i[\Phi,(\chi^1_A
+2\theta\; \tau)A_i]
-\theta\epsilon_{jk}[\nabla_j\Phi,\nabla_kA_i]_+\cr & &\cr
[D_i,\chi^2_A -2\theta\; \tau]\;\Phi &=&\;\qquad   -i[\Phi,(\chi^2_A
-2\theta\; \tau)A_i]
+\theta\epsilon_{jk}[\nabla_j\Phi,\nabla_kA_i]_+\cr & &\cr
[D_i,\chi^3_A ]\;\Phi &=&\nabla_i\Phi   -i[\Phi,\chi^3_A
A_i]\;,\label{fnccdicha}\eeqa where we used (\ref{nbltcmt}). 
The left hand sides of  (\ref{fnccdicha}) can be computed directly.
For this note that  $D_i\Phi$ is covariant, and so the action of
$\chi^a_{A}$  can be simply read off the results 
 (\ref{fncchiaoph}), (\ref{fncc1wt}) and (\ref{fncc2wt})
\beqa
(\chi^1_{A}+2\theta\;\tau)D_i\Phi&=&\;\;\; \frac\theta 2\epsilon_{jk} [ A_j,
D_kD_i\Phi +\nabla_kD_i\Phi ]_+ + H^{(1)}_{D_i\Phi}
\cr& &\cr
(\chi^2_{A}-2\theta\;\tau)D_i\Phi&=& -\frac\theta 2\epsilon_{jk} [ A_j,
D_kD_i\Phi +\nabla_kD_i\Phi ]_+ +
H^{(2)}_{D_i\Phi}\cr& &\cr
\chi^3_{A}D_i\Phi&=&\qquad\qquad\qquad\qquad\qquad\qquad\qquad\;\; H^{(3)}_{D_i\Phi}\;,\eeqa 
where the homogenous terms $H_{D_i\Phi}^{(a)}$ satisfy
$H_{D_i\Phi}^{(3)}=H_{D_i\Phi}^{(1)}+H_{D_i\Phi}^{(2)}$. Using this result  along with (\ref{fncchiaoph}), (\ref{fncc1wt}) and
(\ref{fncc2wt}) gives  
\beqa
[D_i,\chi^1_A +2\theta \tau]\Phi &=&\;\;\;
\theta
\epsilon_{jk}\biggl([F_{ij},D_k\Phi]_++[\nabla_jA_i,\nabla_k\Phi]_+-\frac
i2 [\Phi,[A_k,F_{ij}-\nabla_jA_i]_+]\biggr)\cr & &\cr & & +\; D_iH^{(1)}_\Phi -
H^{(1)}_{D_i\Phi}\cr& &\cr[D_i,\chi^2_A -2\theta \tau]\Phi &=&-\theta
\epsilon_{jk}\biggl([F_{ij},D_k\Phi]_++[\nabla_jA_i,\nabla_k\Phi]_+-\frac
i2 [\Phi,[A_k,F_{ij}-\nabla_jA_i]_+]\biggr)\cr & &\cr & &
 +\; D_iH^{(2)}_\Phi -
H^{(2)}_{D_i\Phi}\cr& &\cr[D_i,\chi^3_A ]\Phi &=&\quad
 D_iH_\Phi^{(3)} -H_{D_i\Phi}^{(3)}\label{fnccdtlycdc} \eeqa 
Finally by comparing  (\ref{fnccdicha}) with  (\ref{fnccdtlycdc}) we
obtain the following general solution to $\chi_A^aA_i$
\beqa \chi^1_AA_i &=& -A_i +\frac\theta 2\epsilon_{jk}[A_k,
F_{ij}-\nabla_jA_i]_+ -2\theta\tau(A_i) +H^{(1)}_{A_i}\cr& &\cr
\chi^2_AA_i &=& \qquad-\;\frac\theta 2\epsilon_{jk}[A_k,
F_{ij}-\nabla_jA_i]_+ +2\theta\tau(A_i) +H^{(2)}_{A_i}\cr& &\cr\chi^3_AA_i
&=& -A_i\qquad\qquad\qquad\qquad\qquad\qquad\qquad\quad\;\; +H_{A_i}^{(3)}\;,\label{chiaona} \eeqa where $H_{A_i}^{(a)}$ are  homogeneous terms,
i.e. they are covariant under gauge transformations.  They are in
general undetermined, apart from the following relations:
\beqa H^{(1)}_{D_i\Phi}&=& D_iH^{(1)}_\Phi+\theta\epsilon_{jk}
[F_{ij},D_k\Phi]_+ +i[\Phi,H^{(1)}_{A_i}]-D_i\Phi\cr& &\cr H^{(2)}_{D_i\Phi}&=& D_iH^{(2)}_\Phi-\theta\epsilon_{jk}
[F_{ij},D_k\Phi]_+ +i[\Phi,H^{(2)}_{A_i}]\cr& &\cr  H_{D_i\Phi}^{(3)}&=&
D_iH_\Phi^{(3)}\qquad\qquad\qquad\qquad +i[\Phi,H_{A_i}^{(3)}]-D_i\Phi\;,\label{rlsbtwnhdpaha}\eeqa 
in
addition to  $H_{A_i}^{(3)}=H_{A_i}^{(1)}+H_{A_i}^{(2)}$.   If
   $H_{{ A}_i}^{(3)}$ vanishes, then $ { A}_i$ is an eigenvector of
   $ \chi^3_{ A}$ with eigenvalue $-1$.  (Minus the eigenvalue of    $
   \chi^3_{ A}$     is analogous to the conformal weight in conformal
   field theory.)

Variations $\delta^{s_a}$ of the field strength 
\beqa {\tt i)} \quad \delta^{s_1}_\epsilon F_{ij}
=D_i\delta^{s_1}_\epsilon A_j-D_j\delta^{s_1}_\epsilon A_i
&=&  \epsilon\;\biggl(\chi^1_{ A} F_{ij} -\frac12 [{\bf x}_k,\nabla_k
F_{ij}]_++2\theta\;\tau(F_{ij})\biggr)\; \cr & &\cr
{\tt ii})\quad \delta^{s_2}_\epsilon  F_{ij}=D_i\delta^{s_2}_\epsilon A_j-D_j\delta^{s_2}_\epsilon A_i
&=& \epsilon\;(\chi^2_{ A}F_{ij}
  -2\theta
\nabla_\theta F_{ij} )\;
\cr & &\cr 
{\tt iii)}\quad \delta^{s_3}_\epsilon F_{ij} =D_i\delta^{s_3}_\epsilon A_j-D_j\delta^{s_3}_\epsilon A_i
&=& \epsilon\;(\chi^3_{ A}F_{ij} -DF_{ij})\;\label{fncdfvrsfij}\;\eeqa
are straightforward since $F_{ij}$ is covariant.  The action of
$\chi^a_{A}$  can again be  read off  
 (\ref{fncchiaoph}), (\ref{fncc1wt}) and (\ref{fncc2wt})
\beqa
(\chi^1_{A}+2\theta\;\tau)F_{ij}&=&\;\;\; \frac\theta 2\epsilon_{jk} [ A_j,
D_kF_{ij} +\nabla_kF_{ij} ]_+ + H^{(1)}_{F_{ij}}
\cr& &\cr
(\chi^2_{A}-2\theta\;\tau)F_{ij}&=& -\frac\theta 2\epsilon_{jk} [ A_j,
D_kF_{ij} +\nabla_kF_{ij} ]_+ +
H^{(2)}_{F_{ij}}\cr& &\cr
\chi^3_{A}F_{ij}&=&\qquad\qquad\qquad\qquad\qquad\qquad\quad\; H_{F_{ij}}^{(3)}\eeqa 
Using (\ref{fncdfvrsfij}) some work shows  that the  homogeneous terms $H^{(a)}_{{ F}_{ij}}$ are related to $ H_{{ A}_i}^{(a)}$ by
\beqa
 H^{(1)}_{{ F}_{ij}}& =& -2{ F}_{ij} +{ D}_i  H^{(1)}_{{
   A}_j } - { D}_j  H^{(1)}_{{\cal
   A}_i}-\theta \epsilon_{k\ell}[ { F}_{ik},{ F}_{j\ell}]_+\cr & &\cr
 H^{(2)}_{{ F}_{ij}}& =&\qquad\quad\;\;\;{ D}_i  H^{(2)}_{{   A}_j } - { D}_j
 H^{(2)}_{{   A}_i}+\theta \epsilon_{k\ell}[ { F}_{ik},{ F}_{j\ell}]_+\cr & &\cr
H^{(3)}_{{ F}_{ij}}& =& -2{ F}_{ij} +{ D}_i  H^{(3)}_{{   A}_j } - {D}_j  H^{(3)}_{{
    A}_i}\;,\label{cnhahf}\eeqa and thus satisfy $H^{(3)}_{{ F}_{ij}}= H_{{
    F}_{ij}}^{(1)}+ H_{{   F}_{ij}}^{(2)}$.    If
   $H^{(3)}_{{ A}_i}$ vanishes, then $ { F}_{ij}$ is an eigenvector of
   $ \chi^3_{ A}$ with eigenvalue $-2$.   For Seiberg-Witten
   transformations $H_{A_i}^{(2)}$ is commonly set to zero, implying
$ H^{(2)}_{{ F}_{ij}} =\theta \epsilon_{k\ell}[ { F}_{ik},{ F}_{j\ell}]_+$.

\subsection{One parameter family of Seiberg-Witten transformations}

In section 2.2, we 
considered  a one parameter family of  Kontsevich maps
(\ref{opfkms}) which led to   a one-parameter family of star
products (\ref{opfosps})
connecting the Voros and  Moyal Weyl products.
In that case 
 $\tau_\beta =\frac{\beta}4\;\nabla_i\nabla_i$, and we have a one
 parameter family of Seiberg-Witten variations on the gauge fields and
 gauge parameter
\beqa\delta^{SW}\Phi&=&\delta\theta\biggl( -\frac14\epsilon_{ij} [ A_i,(
D_j +\nabla_j)\Phi ]_++\frac{\beta}4\;\nabla_i\nabla_i\Phi +\frac{1}{2\theta} H^{(2)}_\Phi\biggr)\cr& &\cr
\delta^{SW}A_i &=&\delta\theta\biggl( -\;\frac 14\epsilon_{jk}[A_k,
F_{ij}-\nabla_jA_i]_+ +\frac{\beta}4\;\nabla_j\nabla_jA_i +\frac{1}{2\theta}H^{(2)}_{A_i}\biggr)\cr& &\cr
\delta^{SW}F_{ij}&=&\delta\theta\biggl( -\frac 14\epsilon_{jk} [ A_j,(
D_k +\nabla_k)F_{ij} ]_++  \frac{\beta}4\;\nabla_k\nabla_kF_{ij} +\frac{1}{2\theta}
H^{(2)}_{F_{ij}}\biggr)\;,\cr & &\cr
\delta^{SW}\Lambda & =& \delta\theta \biggl(\frac 14
\epsilon_{ij} [A_i,\nabla_j \Lambda]_+   +
\frac{\beta}4\;\nabla_i\nabla_i\Lambda\biggr)+ \Lambda_\epsilon^{(2)}\label{opfoswms}\eeqa  
  Using
 the identities
\beqa D_iD_i\Phi& =&\nabla_i\nabla_i \Phi -i[A_i, (D_i+\nabla_i)\Phi] +
i[\Phi, \nabla_iA_i]\cr& &\cr D_j F_{ji}&=&\nabla_j\nabla_j
A_i+i[A_j,F_{ij}-\nabla_jA_i]-D_i\nabla_jA_j \;,\eeqa 
the transformations of the gauge fields in (\ref{opfoswms})  can be re-expressed, up to
gauge transformations,  as 
\beqa\delta^{SW}\Phi&=&\delta\theta\biggl( -\frac14\epsilon_{ij} [ A_i,(
D_j +\nabla_j)\Phi ]_+   +\frac{i\beta}4\;[A_i, (D_i+\nabla_i)\Phi] +\frac{1}{2\theta}\tilde H^{(2)}_\Phi\biggr)\cr& &\cr
\delta^{SW}A_i &=&\delta\theta\biggl( -\;\frac 14\epsilon_{jk}[A_k,
F_{ij}-\nabla_jA_i]_+ -\frac{i\beta}4\;[A_j,F_{ij}-\nabla_jA_i]
+\frac{1}{2\theta}\tilde H^{(2)}_{A_i}\biggr)\cr& &\cr
\delta^{SW}F_{ij}&=&\delta\theta\biggl( -\frac 14\epsilon_{jk} [ A_j,(
D_k +\nabla_k)F_{ij} ]_++  \frac{i\beta}4\;[A_k, (D_k+\nabla_k)F_{ij}] +\frac{1}{2\theta}\tilde
H^{(2)}_{F_{ij}}\biggr)\;,\label{thrtwntsvn}
\eeqa where we redefined the homogeneous terms: \beqa  \tilde H^{(2)}_\Phi&=& H^{(2)}_\Phi+ \frac{\beta\theta}2\;
D_iD_i\Phi\cr & &\cr  \tilde H^{(2)}_{A_i}&=& H^{(2)}_{A_i}+\frac{\beta\theta}2 D_jF_{ji}\cr& &\cr  \tilde H^{(2)}_{F_{ij}}&=& H^{(2)}_{F_{ij}}+
\frac{\beta\theta}2\; D_kD_k F_{ij}\eeqa  For $\beta=1$ we get the Seiberg-Witten map for the Voros star
 product, while for $\beta=0$ it is  the Seiberg-Witten map for the Moyal Weyl star product.  The $\beta-$dependent inhomogeneous
terms  in   (\ref{thrtwntsvn}) are expressed in terms of
commutators, which then vanish at lowest order in $\theta$.  From
(\ref{thrtwntsvn}), 
 $\tau_\beta$ acting on  fields belonging to nontrivial representations of
 the gauge group has the correct $\theta\rightarrow 0$ limit, which was
 not obvious from  (\ref{opfoswms}).
 In  going from  (\ref{opfoswms}) to (\ref{thrtwntsvn}) we have  absorbed
 the lowest order  $\theta$ contributions to the $\beta-$dependent
 inhomogeneous  terms of  (\ref{opfoswms})
 into the homogeneous terms. 
The $\theta\rightarrow 0$ limit  is then  in agreement with (\ref{fostswes}), and  consistent with the fact
the first order result for the Seiberg-Witten map should not depend on the choice of star product.

\subsection{Covariant position operator}

For gauge  theories on the noncommutative plane one can define a
covariant position operator $X_i$:
\be X_i =\frac{\hat {\tt x}_i}\theta + \epsilon_{ij} A_j\;,  \ee
where $\hat {\tt x}_i$ are gauge invariant functions such that $\hat
{\tt x}_i({\tt x},\theta)={\tt x}_i$.  One can obtain the variations
$\delta^{s_a}_\epsilon X_i$ by computing the action  of $\chi^a_{A}$.
The Seiberg-Witten variations  $\delta^{SW}X_i$ were previously
obtained in \cite{Polychronakos:2002pm} for the Moyal Weyl case $\tau=0$.
Since $X_i$ is covariant,   the action of $\chi^a_{A}$  on it
 can again be  read off  
 (\ref{fncchiaoph}), (\ref{fncc1wt}) and (\ref{fncc2wt})
\beqa 
(\chi^1_{A}+2\theta\;\tau)X_{i}&=&-2\epsilon_{ij} A_j + \frac\theta 2
\biggl( [ A_j,
F_{ji} -\nabla_i A_{j} ]_+ + [A_i,\nabla_jA_j]_+\biggr) + H^{(1)}_{X_{i}}
\cr& &\cr
(\chi^2_{A}-2\theta\;\tau)X_{i}&=&\;\;\;2\epsilon_{ij} A_j - \frac\theta 2
\biggl( [ A_j,
F_{ji} -\nabla_i A_{j} ]_+ + [A_i,\nabla_jA_j]_+\biggr) + H^{(2)}_{X_{i}}
\cr& &\cr
\chi^3_{A}X_{i}&=&\qquad\qquad\qquad\qquad\qquad\qquad\qquad\qquad\qquad\qquad\qquad\; H_{X_{i}}^{(3)}\eeqa 
 where $H_{X_i}^{(a)}$ are  homogeneous terms, satisfying
  $H_{X_i}^{(3)}=H_{X_i}^{(1)}+H_{X_i}^{(2)}$.  Using (\ref{chiaona}), we can
  compare this with $\chi^a_{A}$ 
acting on $\epsilon_{ij} A_j$, and then deduce the action of
$\;\chi^a_{A}$  on the ratio of the noncommutative coordinate with $\theta$.  Up to homogeneous terms, one gets
\beqa
(\chi^1_{A}+2\theta\;\tau)\;\frac{\hat {\tt
    x}_i}\theta&=&\;\;\;\;\;\frac{\hat {\tt
    x}_i}\theta\cr & &\cr(\chi^2_{A}-2\theta\;\tau)\;\frac{\hat {\tt
    x}_i}\theta&=&-2\;\frac{\hat {\tt
    x}_i}\theta\cr & &\cr\chi^3_{A}\;\frac{\hat {\tt
    x}_i}\theta&=&-\;\;\;\frac{\hat {\tt
    x}_i}\theta\label{chonxot}\eeqa
Because 
 variations $\delta^{s_a}_\epsilon$ are evaluated at  fixed coordinates on
 the noncommutative plane and fixed values of $\theta$, the variations
 of $\hat{\tt x}_i$ and $\theta$, and consequently their ratio, should
 vanish. This follows if (\ref{chonxot}) is an exact result, {\it including
 the homogeneous terms}.  As a result the homogeneous terms
$H^{(a)}_{X_{i}}$ and $H^{(a)}_{A_{i}}$ are related by
\beqa H^{(1)}_{X_{i}}&=&\epsilon_{ij}H^{(1)}_{A_{j}} +\;X_i\cr & &\cr H^{(2)}_{X_{i}}&=&\epsilon_{ij}H^{(2)}_{A_{j}} -2X_i\cr & &\cr H^{(3)}_{X_{i}}&=&\epsilon_{ij}H^{(3)}_{A_{j}} -\;\;X_i \label{rbhxaha}\eeqa

\subsection{Constraints on the Homogeneous Terms} 

In the above we have found a number of relations connecting the various
homogeneous terms: (\ref{rlsbtwnhdpaha}) and
(\ref{cnhahf}).  From (\ref{rlsbtwnhdpaha}) it
follows that not all terms $H^{(a)}_{ { D}_i\Phi}$, $ H^{(a)}_\Phi$ and $H^{(a)}_{{
    A}_i}$, for any $a$, can simultaneously vanish.  From
(\ref{cnhahf}), $H^{(a)}_{{
    A}_i}$ and $H^{(a)}_{ F_{ij}}$, for any $a$, cannot
simultaneously vanish.  We can get additional
 constraints on the homogenous terms  if we demand that 
the gauge fields carry a faithful representation of the two independent scale transformations  {\tt i)} and
{\tt  ii)}.   When including this  demand   there appear to be no obstructions to having 
   nonvanishing homogeneous terms.   We give some explicit solutions.
   The new conditions are however insufficient in removing all the
ambiguities in the homogeneous terms.   More constraints may result from the presence
of other symmetries (although they won't be considered here) and they may help fix further 
degrees of freedom in the homogeneous terms.

   At first order, the   two independent scale transformations  {\tt i)} and
{\tt  ii)} transformations commute.  An
analogous statement can be made at all orders using the scale
generators.   Acting   on an arbitrary function $F$, they are
$\frac 12[{\tt   x}_i,\nabla_iF]_+$ and $2\theta
(\nabla_\theta -\tau)F$.  Their commutator acting on $F$
vanishes, \beqa & & \frac 12[{\tt   x}_i,\nabla_i\;2\theta
(\nabla_\theta -\tau)F]_+ - 2\theta
(\nabla_\theta -\tau)\;\frac 12[{\tt   x}_i,\nabla_iF]_+ 
&\cr & &\cr &=&\frac 12\biggl([{\tt   x}_i,\nabla_i\;DF]_+ - D[{\tt   x}_i,\nabla_iF]_+\biggr)\cr & &\cr &=&\frac 12\biggl([{\tt   x}_i,[\nabla_i,D]F]_+ - [D{\tt   x}_i,\nabla_iF]_+\biggr)
\;=\; 0 \;, \eeqa where we used (\ref{txice}).  So for gauge fields to
carry a faithful representation we need that  {\tt i)} and
{\tt  ii)} commute on the space of such fields.    (More generally,
one only needs that the commutator of  {\tt i)} and
{\tt  ii)} is a gauge transformation.) 

   We now compute $ [\delta^{s_1}_{\epsilon_1},
\delta^{s_2}_{\epsilon_2}]$.   Acting on $\Phi$ we get 
\beqa  [\delta^{s_1}_{\epsilon_1}, \delta^{s_2}_{\epsilon_2}]\;\Phi &=&
[\delta^{s_3}_{\epsilon_1}, \delta^{s_2}_{\epsilon_2}]\;\Phi=\epsilon_1\epsilon_2\;\biggl\{\chi^3_{ A}H^{(2)}_\Phi-(\chi^2_{
A}-2\theta\;\tau) H^{(3)}_\Phi\\
& &\cr &-&\frac\theta 2\epsilon_{ij}\;\biggl( [ A_i,  ( D_j +
\nabla_j)H^{(3)}_\Phi + i[\Phi, H^{(3)}_{A_j}]\; ]_+ + [ H^{(3)}_{ A_i},  ({ D}_j + \nabla_j)\Phi]_+ \biggr)
\biggr\}\nonumber \eeqa  An obvious solution to $ [\delta^{s_1}_{\epsilon_1}, \delta^{s_2}_{\epsilon_2}]\;\Phi=0$ is $ H^{(3)}_{{
    A}_i}=H^{(a)}_\Phi=0\;,\;\;a=1,2,3$.  More general solutions are also
possible.   For example, setting $ H^{(3)}_{{
    A}_i}=0$, and  $H^{(2)}_\Phi$ and $H^{(3)}_\Phi$ equal to
functions  only of $\Phi$ 
  we need that \be H^{(2)}_\Phi\; \frac\delta{\delta\Phi}
\; H^{(3)}_\Phi =H^{(3)}_\Phi\; \frac\delta{\delta\Phi}
\; H^{(2)}_\Phi\;,\label{slnfraiez} \ee   using (\ref{fncchiaoph}) and (\ref{fncc2wt}).
(\ref{slnfraiez})  is
 then solved for $H^{(3)}_\Phi$, $H^{(2)}_\Phi$, and consequently  $H^{(1)}_\Phi$,  proportional to the same function of $\Phi$.

Acting on ${ A}_i$ the commutator gives 
\beqa  [\delta^{s_1}_{\epsilon_1}, \delta^{s_2}_{\epsilon_2}]\;{ A}_i &=&
[\delta^{s_3}_{\epsilon_1}, \delta^{s_2}_{\epsilon_2}]\;{ A}_i\label{cmtracnai}\\
& &\cr &=&\epsilon_1\epsilon_2\;\biggl\{(\chi^3_{{ A}}+1)H^{(2)}_{{    A}_i}-(\chi^2_{{ A}}\;-\;2\theta\;\tau)H^{(3)}_{{ A}_i}
\cr & &\qquad-\frac\theta 2\epsilon_{jk}\;\biggl([ { A}_k,D_i H^{(3)}_{A_j}
- D_j H^{(3)}_{A_j} -
\nabla_j   H^{(3)}_{   A_i }]_++ [  H^{(3)}_{{   A}_k}  ,   F_{ij}-\nabla_j A_i]_+\biggr)\biggr\}\;,
\nonumber \eeqa where we used (\ref{cnhahf}).  $ [\delta^{s_1}_{\epsilon_1}, \delta^{s_2}_{\epsilon_2}]\;{ A}_i=0 $ is obviously
satisfied for $ H^{(a)}_{{ A}_i}=0\;,\;\;a=1,2,3\;$.
It is not difficult to construct other
solutions to $ [\delta^{s_1}_{\epsilon_1},
\delta^{s_2}_{\epsilon_2}]\;{ A}_i=0 $.   For example,
if there is a covariant vector  $V_i$ which is an eigenvector of
$\chi^3_{ A}$ with eigenvalue $-1$, we can satisfy (\ref{cmtracnai}) by
setting 
 \be  H^{(1)}_{{ A}_i}=- H^{(2)}_{{ A}_i} =V_i\;,\qquad H^{(3)}_{{
     A}_i}=0\ee    For the case of pure gauge theory, we can let $V_i$
  be proportional to the covariant position operator $X_i$.  It has 
eigenvalue $-1$, since from (\ref{rbhxaha}), $ H^{(3)}_{X_{i}}= -X_i$.
When other fields are present, $V_i$ can have additional
contributions.  For example in scalar field
  theory such a contribution can be chosen to be proportional to $ {
    D}_i \Phi$ provided that $H^{(3)}_\Phi=0$, and using (\ref{fnccdtlycdc}).

\section{Concluding Remarks}
\setcounter{equation}{0}

 We have seen that one of the
 previous transformations, namely {\tt iii)},
preserves the fundamental commutation relations.  It was generated by
 $D$ which satisfies the Leibniz rule, as well as  annihilates the left hand side of
(\ref{fndmtlcrs}), resulting in a symmetry of the algebra.
Here we remark on the possibility of implementing  transformation {\tt iii)}
  as a  symmetry of the dynamics as well.  We note below that this  symmetry
is not 
a deformation of the standard dilation symmetry on the (commutative)  plane, but
 rather a new symmetry on the noncommutative plane $\times\;
 {\mathbb{R}}$.   For simplicity, we shall restrict the
 remarks here to lowest order in $\theta$.   We plan to give a more thorough
 discussion on this topic in a later article. 

Since {\tt iii)} involves re-scalings in both $x_i$ and
$\theta$, we cannot keep $\theta$ fixed. So  we should not consider a
single noncommutative plane, but rather an ensemble of such noncommutative
planes.  $\theta$ characterizing the noncommutative plane is often
related to some external field. (In string theory it is associated
with the external 2-form on the brane, while in the Hall effect it is
the external magnetic field.)  From that point of view we are then 
allowing for changing values of the field.  Let us assume the changes
occur in some (commuting) time variable  $t$, i.e. the field and
therefore $\theta$ are  functions of the time $t$.  From a
cosmological perspective, it is tempting to imagine a $\theta(t)$  where $\theta$ monotamicaly
decreases with increasing $t$, for then one would have a model where
noncommutativity was significant  at  early times, while it is
negligible for late
(or present) times.  

At first order we write the action $S$  as an integral of the
Lagrangian density $L(x,t)$ over
${\mathbb{R}}^3$, here
parametrized by  $(x_1,\;x_2,t)$
\be S=\int dt d^2x\;  L(x,t)  \ee  If we assume that $\theta(t)$  is a
nonsingular function, we can do a change of variables
\be S=\int d\theta d^2x\; {\cal L}(x,\theta) \;,\qquad L(x,t)=
\bigg|\frac{d \theta}{dt}\bigg|\;{\cal L}(x,\theta) \ee  This action is
invariant under ${\tt iii)}$ when the Lagrangian density ${\cal
  L}(x,\theta)$ transforms as \be {\cal L}(x,\theta) \rightarrow 
{}^{\rho_3} {\cal L} (x,\theta)
= \rho^4\;{\cal L}(\rho x,\rho^2\theta)\ee
The corresponding infinitesimal transformation is
\be {\cal L} (x,\theta) \rightarrow {\cal L} (x,\theta)
+ \delta^{s_3}_\epsilon {\cal L} (x,\theta)  \;,\ee where variation
of the Lagrangian density is a total divergence in ${\mathbb{R}}^3$,
$$  \delta^{s_3}_\epsilon {\cal L}(x,\theta)
\;=\; -\epsilon\;(4+D){\cal L}(x,\theta)\;=\; \partial_i C^i+
\partial_\theta C^\theta\;,$$
 \be C^i = -\epsilon x_i{\cal L}\qquad C^\theta =
-2\epsilon\theta {\cal L}\ee
 In terms of the previous notation
$\chi^3_{\cal A}{\cal L}=-4\;{\cal L}\;$, or ${\cal L}$ has conformal
weight four.    It is easy to write  a
Lagrangian field theory with this property, as Lagrangian densities
associated  with {\it four} dimensional
 conformal symmetry  transform in the same
 manner under dilations.  An example is  
scalar field theory with a $\phi^4$ interaction.  At zeroth order in
$\theta$, the Lagrangian density is 
\be {\cal L} =\frac 12 (\partial_i \phi)^2+ \frac
g{4!}\phi^4\;,\label{pttfthlgrngn}\ee where $g$ is a dimensionless
coupling and  the scalar field $\phi(x,\theta)$ satisfies 
$\chi^3_{\cal A}\phi=-\;\phi\;,$ or \be  \delta^{s_3}_\epsilon\phi(x,\theta)
= -\epsilon\;(1+D)\phi(x,\theta)\;\label{dstfsf}\ee Thus the conformal
weight of $\phi(x,\theta)$ is one, in agreement with scalar fields in four
dimensional conformal field theory.    As in four
dimensions, the mass term
breaks the scale symmetry.

From Noether's theorem the above symmetry implies a conserved current,
but it is a conserved current in  ${\mathbb{R}}^3$.  The current is
not conserved on two
dimensional $\theta$-slices of  ${\mathbb{R}}^3$.   So denoting the
current as $j^\mu$,
$\mu=1,2,\theta$, the conservation law is 
\be  \partial_i j^i+
\partial_\theta j^\theta=0\ee For generic  fields $\psi_\alpha$ the current
is given by
\be \epsilon j^\mu = \frac{\partial{\cal L}}{\partial\partial_\mu \psi_\alpha} 
 \delta^{s_3}_\epsilon\psi_\alpha -C^\mu\ee  In the example of the scalar field
 $\phi$ with Lagrangian density (\ref{pttfthlgrngn})  we then get 
\be j^i = -\partial_i\phi (1+D)\phi +\frac 12 x_i (\partial_j \phi)^2+
\frac g{4!}x_i \phi^4\qquad
\qquad j^\theta = 2\theta\;\biggl(\frac 12 (\partial_j \phi)^2+ \frac g{4!}\phi^4\biggr) \ee
We note that although $j^\theta$ vanishes in the commutative limit
$\theta\rightarrow 0$, it nevertheless gives a nonzero contribution to the
divergence when $\theta\rightarrow 0$, and so the
current is not conserved when restricted to the
$\theta= 0 $ slice of ${\mathbb{R}}^3$.  For the example of the scalar
field,
$\partial_i j^i|_{\theta=0}= -2\;{\cal L}$.   (More generally, this
holds provided ${\cal L}$ doesn't depend on $\theta$-derivatives of
the fields.)   The  symmetry
transformation  then cannot be regarded as
a deformation of the standard dilation symmetry on the plane.  For
 the free scalar
field, the latter is associated with variations   $\delta^{s_1}_\epsilon
\phi=- \epsilon\; x_i\partial_i\phi\;$, corresponding to conformal
weight zero, which differs from the
$\theta\rightarrow 0 $ limit of variations (\ref{dstfsf}),  $\delta^{s_3}_\epsilon
\phi|_{\theta=0}=- \epsilon\;(1+ x_i\partial_i)\phi\;$, having  conformal
weight one.

\bigskip
\bigskip

{\parindent 0cm{\bf Acknowledgement}}                                         
 
We are very grateful to J.~M.~Grimstrup, D. O'Connor and
P. Pre\v{s}najder for  useful 
discussions.  This work was supported in part by the joint NSF-CONACyT grant
E120.0462/2000 and  DOE grants   DE-FG02-85ER40231 and DE-FG02-96ER40967.

\end{document}